\documentclass[twocolumn,trackchanges]{aastex701}

\usepackage{CJKutf8}
\newcommand{\namecn}[1]{\begin{CJK*}{UTF8}{gbsn}({#1})\end{CJK*}}
\usepackage{amsmath}
\usepackage{amssymb}
\usepackage{rotating}

\begin{document}

\title{Transiting Planetary Systems with Distant Giant Companions Remain Moderately Coplanar}


\correspondingauthor{Fabo Feng}
\email{ffeng@sjtu.edu.cn}

\author[orcid=0000-0001-6753-4611]{Guang-Yao Xiao \namecn{肖光耀}}
\affiliation{State Key Laboratory of Dark Matter Physics, Tsung-Dao Lee Institute \& School of Physics and Astronomy, Shanghai Jiao Tong University, Shanghai 201210, China}
\email{gyxiao\_tdli@sjtu.edu.cn}

\author[orcid=0000-0003-3860-6297]{Huan-Yu Teng \namecn{滕环宇}}
\affiliation{CAS Key Laboratory of Optical Astronomy, National Astronomical Observatories, Chinese Academy of Sciences, Beijing 100101, China}
\email{hyteng@bao.ac.cn}

\author[orcid=0000-0002-2546-2012]{Xiumin Huang \namecn{黄秀敏}}
\affiliation{State Key Laboratory of Dark Matter Physics, Tsung-Dao Lee Institute \& School of Physics and Astronomy, Shanghai Jiao Tong University, Shanghai 201210, China}
\email{xm_huang@sjtu.edu.cn}

\author[orcid=0000-0001-6039-0555]{Fabo Feng \namecn{冯发波}}
\affiliation{State Key Laboratory of Dark Matter Physics, Tsung-Dao Lee Institute \& School of Physics and Astronomy, Shanghai Jiao Tong University, Shanghai 201210, China}
\email{ffeng@sjtu.edu.cn}

\author[orcid=0000-0002-1934-6250]{Dong Lai \namecn{赖东}}
\affiliation{State Key Laboratory of Dark Matter Physics, Tsung-Dao Lee Institute \& School of Physics and Astronomy, Shanghai Jiao Tong University, Shanghai 201210, China}
\email{donglai@sjtu.edu.cn}

\author[orcid=0000-0002-8958-0683]{Fei Dai \namecn{戴飞}}
\affiliation{Institute for Astronomy, University of Hawai`i, 2680 Woodlawn Drive, Honolulu, HI 96822 USA}
\email{fdai@hawaii.edu}

\author[0009-0008-3430-1027]{Yu-Juan Liu\namecn{刘玉娟}}
\affiliation{CAS Key Laboratory of Optical Astronomy, National Astronomical Observatories, Chinese Academy of Sciences, Beijing 100101, China}
\email{lyj@bao.ac.cn}

\begin{abstract}
The mutual inclination between inner planets and distant giant companions provides an important probe of planetary system formation and dynamical evolution, yet direct measurements of this quantity remain scarce. We combine radial velocity (RV) observations with Hipparcos--Gaia astrometry to constrain the orbital architecture of 19 planetary systems hosting at least one transiting inner planet and one outer giant companion. Using a hierarchical Bayesian framework, we infer the population-level distribution of the minimum mutual inclination, $\Delta I$, between the inner and outer planetary orbits.
We find that the $\Delta I$ distribution is well described by a Rayleigh model with a scale parameter of $\sigma = 15.8^{+2.8}_{-2.6}\deg$, which is strongly preferred over an isotropic distribution ($\Delta\log Z=5.45$). This result suggests that transiting systems hosting distant giant companions remain substantially more coplanar than expected for an isotropic population, consistent with the partial preservation of primordial coplanarity.
A division by the mass ($0.3\,M_{\rm Jup}$) of the inner transiting planet suggests that giant-inner-planet systems may have lower $\Delta I$ than small-inner-planet systems, with $P(\sigma_{\rm giant}<\sigma_{\rm small})=0.952$; however, the current data do not significantly favor a model allowing different $\sigma$ values for the two subsamples over one in which they share a common $\sigma$.
Future Gaia DR4 astrometry will enable more robust population-level studies of the three-dimensional architectures of systems with distant giant companions.
\end{abstract}


\section{Introduction}
Planetary systems are believed to form within a flat protoplanetary disk, naturally producing nearly coplanar orbital architectures. The Solar System provides a typical example, with the major planets orbiting within only a few degrees of a common plane. Similar alignments have been observed in several directly imaged planetary systems. For example, the orbital plane of $\beta$ Pictoris b is closely aligned with the debris disk surrounding the host star \citep{Lagrange2019NatAs}, while the two giant planets in the young PDS~70 system appear broadly coplanar with the protoplanetary disk from which they are still accreting material \citep{Wang2021AJ}. These observations suggest that coplanarity is a common outcome of planet formation and motivate efforts to measure the three-dimensional (3D) architectures of planetary systems.

A fundamental quantity describing planetary-system architecture is the mutual inclination, defined as the angle between the orbital angular momentum vectors of two planets. This quantity provides a sensitive probe of both formation and subsequent dynamical evolution, as processes such as planet--planet scattering, secular interactions, resonant migration, and perturbations from distant companions can alter orbital inclinations over time \citep{Chatterjee2008ApJ,FabryckyTremaine2007ApJ,Naoz2016ARA&A,Dawson2018ARA&A}. However, direct measurements of mutual inclination remain challenging, especially in non-transiting systems due to limited information on orbital orientation, e.g., owing to the lack of constraints on the inclination ($I$) and longitude of ascending node ($\Omega$). To date, only a handful of systems possess well-constrained true mutual inclination from observation. Roughly half of them are determined via dynamical modeling of transit timing and/or duration variations (TTVs and TDVs), e.g., Kepler-448 \citep{Masuda2017AJ}. The remainder are mainly obtained from the joint analysis of multiple techniques, including RV, Hipparcos-Gaia astrometry, and ground- or space-based relative astrometry, e.g., 14 Her \citep{Benedict2023AJ, Bardalez2025ApJ, Xiao2025RNAAS}.

Although the true mutual inclination remains inaccessible in most cases, transiting systems permit measurements of the minimum mutual inclination, $\Delta I=|I_{\rm in}-I_{\rm out}|$, which represents a lower bound on the true mutual inclination. Because transiting planets have orbital inclinations close to $90^\circ$, the problem simplifies to $\Delta I\approx|90^\circ-I_{\rm out}|$. Consequently, measuring $\Delta I$ requires only a constraint on the inclination of the outer companion, $I_{\rm out}$, which can be obtained from joint RV and astrometric analyses. Minimum mutual inclinations therefore provide a practical observational window into the orbital architectures of planetary systems.

Recent joint analysis of RV and Hipparcos--Gaia astrometry have enabled $\Delta I$ measurements for a growing number of systems hosting transiting planets and outer giant companions. Examples include HAT-P-11 \citep{Xuan2020MNRAS_pimen, An2025AJ_HAT-P-11}, $\pi$ Men \citep{Xuan2020MNRAS_pimen}, HD~118203 \citep{Zhang2024AJ_HD118203}, and HD~73344 \citep{Zhang2025AJ_HD73344}. These systems span a wide range of architectures, from nearly coplanar configurations to substantially inclined systems, implying considerable diversity among planetary systems hosting distant giant companions.

At the population level, statistical analyses of \textit{Kepler} systems have shown that compact multi-planet systems are typically highly coplanar, with characteristic mutual inclinations of only a few degrees \citep{Lissauer2011ApJS,Fang2012ApJ,Dai2018ApJ}. In contrast, \citet{Masuda2020AJ} inferred a broader distribution for systems hosting cold Jupiters, finding characteristic mutual inclinations of order $10^\circ$. More recently, studies of S-type transiting planets in binaries have begun to constrain the relative alignment between planetary and stellar companion orbits \citep{Dupuy2022MNRAS, Christian2022AJ, Lester2023AJ, Rice2024AJ, Christian2025AJ, Zhang2026AJ}. Nevertheless, most existing studies rely either on indirect statistical inference or on stellar-binary architectures, and direct population-level constraints on $\Delta I$ between transiting planets and distant giant planets remain scarce.

In this work, we combine RVs with Hipparcos--Gaia astrometry to constrain the orbital inclinations of outer giant companions in 19 systems hosting at least one transiting inner planet.
We also derive posterior distributions of the $\Delta I$ for individual systems and employ hierarchical Bayesian modeling to infer the underlying population distribution. This approach provides one of the first population-level measurements of the $\Delta I$ distribution in transiting planetary systems hosting distant giant companions.

\section{Sample Selection}
The goal of this work is to investigate the orbital architectures and $\Delta I$ distribution of transiting planetary systems hosting distant outer perturbers. We first constructed a parent sample using the NASA Exoplanet Archive \citep{Christiansen2025} and the \textit{Encyclopaedia of Exoplanetary Systems} \citep{Schneider2011}. The selection criteria were as follows:
\begin{enumerate}
    \item each system must host at least one inner transiting planet together with an outer giant companion having orbital period $P_{\rm out}>400$ days and minimum mass $M_{\rm out}\sin I>0.3\,M_{\rm Jup}$ (or $95\,M_\oplus$);
    
    \item the minimum expected astrometric signature induced by the outer companion must satisfy $\alpha_{\rm out}>0.02$ mas, ensuring that the orbital inclination can potentially be constrained through joint RV and astrometric modeling;
    
    \item systems are restricted to distances within 500 pc.
\end{enumerate}

We exclude companions flagged as controversial in either catalog and further require that the outer companions have been robustly confirmed through RV observations. In addition, after cross-matching with the TESS project candidates, we included two systems, HD\,50554 and HIP\,54597, hosting candidate inner transiting planets that satisfy the above criteria. The inner transiting planets in HD\,50554 were validated by \citet{Liu2026arXiv}.

This procedure yields a parent sample of 35 systems (see Table~\ref{Tab:star} of Appendix~\ref{app:stellar_pars}). The host stars are predominantly FGK main-sequence stars, with stellar masses in the range $0.7\lesssim M_\star/M_\odot\lesssim1.3$ and effective temperatures between $4500\,{\rm K}\lesssim T_{\rm eff}\lesssim6300\,{\rm K}$. The final sample is established using the results of our joint RV+astrometric orbital fits (see Section~\ref{sec:orbit_fit}). We retain only systems with well-constrained inclinations for the outer companion, corresponding to a $1\sigma$ uncertainty smaller than $35^\circ$. This threshold is chosen because systems with larger uncertainties are generally dominated by the inclination prior and do not represent meaningful astrometric detections.
Applying this criterion leaves 19 systems with reliable inclination measurements. According to the property of the innermost transiting planet, the final sample consists of 7 systems hosting super-Earths ($M_{\rm in}\lesssim10\,M_\oplus$), 5 systems hosting Neptune/Saturn-mass planets ($10\lesssim M_{\rm in}/M_\oplus\lesssim95$), and 7 systems hosting Jupiter-mass planets ($M_{\rm in}\gtrsim95\,M_\oplus$).

 
\section{Data}
\subsection{Radial velocity data}
We compile the RV data for each target from the literature, primarily from the original discovery papers and subsequent follow-up observations. Several long-term RV surveys based on high-resolution spectroscopy have also contributed a large number of high-precision measurements. Table~\ref{Tab:RV} of Appendix~\ref{app:rv_data} lists the details of instrument, observation count, time span, and mean RV uncertainty for each star.

\subsection{Absolute astrometric data}
To derive astrometric constraints, we incorporate the absolute astrometry from the new Hipparcos reduction \citep{vanLeeuwen2007} and the Gaia second and third data releases (GDR2 and GDR3; \citealt{GaiaCollaboration2018,GaiaCollaboration2023}), including positions in right ascension ($\alpha$) and declination ($\delta$), proper motions ($\mu_{\alpha}$ and $\mu_{\delta}$), and parallax ($\varpi$). We further include the Hipparcos intermediate astrometric data (IAD; also referred to as abscissa data; \citealt{vanLeeuwen2007}) and synthetic Gaia epoch observations generated using the Gaia Observation Forecast Tool\footnote{\url{https://gaia.esac.esa.int/gost/index.jsp}}(GOST), and perform a joint analysis together with the RV data.

The Hipparcos IAD and Gaia GOST data primarily provide the satellite scan angle $\psi$, the along-scan (AL) parallax factor $f^{\rm AL}$, and the corresponding barycentric observation epochs. Since Gaia epoch astrometry is not publicly available, we use GOST to reconstruct the Gaia observing cadence and scan geometry.

The astrometric anomaly between Hipparcos and Gaia, corresponding to a temporal baseline of $\sim25$ yr, is primarily sensitive to long-period companions \citep{Brandt2018, Brandt2021, Kervella2022, Feng2024ApJS}, whereas the shorter-term anomaly between GDR2 and GDR3 is more sensitive to companions on orbits with periods of order $\sim10^3$ days \citep{Xiao2026ApJ}.

\section{orbit analysis and results}
\label{sec:orbit_fit}
Transiting systems hosting distant RV-detected companions are particularly well suited for our method of constraining orbital architectures. First, the orbital geometry of the inner transiting planet is already well determined from transit observations alone, implying $I_{\rm in}\sim90^\circ$. Second, distant outer perturbers are more likely to produce detectable astrometric signals, enabling additional constraints from astrometric measurements. These advantages substantially simplify the orbital modeling procedure. 

\subsection{Joint analysis of RV and astrometry}
We adopt a right-handed coordinate system defined by the orthonormal triad $[\hat{x}, \hat{y}, \hat{z}]$, where $\hat{x}$ points the direction of increasing R.A., $\hat{y}$ points the direction of increasing decl., and $\hat{z}$ points away from the observer (convention $\mathrm{III}$ of \citealt{Feng2019ApJS..242...25F} or \citealt{Feng2026enap}). This convention is used in the astrometry models of Hipparcos and Gaia \citep{ESA1997, Lindegren2012A&A}. 

We primarily constrain the 3D orbit of the outer companion through a joint fit to the RV and Hipparcos-Gaia astrometric data, following the framework developed by \citet{Feng2023MNRAS} and subsequently modified by \citet{Xiao2024MNRAS}. The technique has been widely used to detect and characterize long-period companions (e.g., \citealt{cui2026PNAS,Wuyiyun2026AJ,Pablo2026A&A,Sang2026RAA}). Compared with the distant outer companions, the inner short-period planets contribute negligible astrometric signals and are therefore ignored in the astrometric modeling. Their dynamical contributions are included only in the RV component of the joint fit.
The primary fitted parameters involved in our model include seven elements (orbital period $P$, RV semi-amplitude $K$, eccentricity $e$, argument of periastron $\omega$ of stellar reflex motion, inclination $I$, longitude of ascending node $\Omega$, mean anomaly $M_{0}$ at J2017.5), five astrometric offsets ($\Delta \alpha*$, $\Delta \delta$, $\Delta \varpi$, $\Delta \mu_{\alpha*}$ and $\Delta \mu_\delta$) of barycenter relative to GDR3, RV jitter terms $\sigma_{\rm jit}$ for each instrument, astrometric jitter $J_{\rm hip}$ for Hipparcos IAD, and error inflation factor $S_{\rm gaia}$ for Gaia astrometry. The semi-major axis $a$ of the planet relative to the host, the mass of planet $M_{\rm p}$, and the epoch of periastron passage $T_{\rm p}$ can be derived from above orbital elements. 
The stellar mass is adopted from the literature and incorporated as a Gaussian distribution when deriving $a$ and $M_{\rm p}$, but is not treated as a free parameter in the orbital fit.

For RV model, the likelihood can be calculated by
\begin{equation}
\begin{split}
  \mathcal{L}_{\rm RV}
  &=\prod\limits_{j=1}^{N_{\rm RV}}\prod\limits_{k=1}^{N_{\rm inst}}
    \frac{1}{\sqrt{2\pi(\sigma_{j,k}^2+\sigma_{{\rm jit},k}^2)}}\times\\
  &\quad\exp\!\left(-\frac{(v_{j,k}-\hat{v}_{j,k}-\gamma_k)^2}
    {2(\sigma_{j,k}^2+\sigma_{{\rm jit},k}^2)}\right)~,
\end{split}
\end{equation}
where $N_{\rm RV}$ and $N_{\rm inst}$ are the number of RV measurements and instruments, and $\gamma_k$ and $\sigma_{{\rm jit},k}$ are the RV zero-point and the jitter term. $\hat{v}_{j,k}$ and $v_{j,k}$ are the predicted and observed RVs, respectively. We have marginalized out the RV zero-point following \citet{Brandt2021borvara}. 

The likelihood for Hipparcos IAD is expressed as
\begin{equation}
  \mathcal{L}_{\rm hip}=\prod\limits_{j=1}^{N_{\rm IAD}}\frac{1}{\sqrt{2\pi(\sigma_{j}^2+J_{\rm hip}^2)}}\times{\rm exp}\left(-\frac{(\hat{\xi}_j-\xi_j)^2}{2(\sigma_{j}^2+J_{\rm hip}^2)}    \right)~,
\end{equation}
where $N_{\rm IAD}$ is the total number of Hipparcos IAD, $\sigma_j$ is the individual measurement uncertainty, and $J_{\rm hip}$ is the jitter term. Predicted and observed 1D abscissae data are denoted with $\hat{\xi}_j$ and $\xi_j$, respectively. The abscissae $\hat{\xi}_j$ is synthesized by projecting the combined motion of the system's barycenter, the stellar reflex motion, and the parallax motion onto the 1D scan direction of Hipparcos. In addition, the offset between Hipparcos astrometry and the astrometry propagated from the GDR3 epoch to the Hipparcos epoch are also incorporated into the modeling of $\hat{\xi}_j$.

The likelihood for GDR2 and GDR3 can be written as
\begin{equation}
\begin{split}
  \mathcal{L}_{\rm gaia}=
  &\prod\limits_{k=1}^{N_{\rm DR}}\frac{1}{\sqrt{(2\pi)^5|\Sigma_k(S^2)|}}\times\\
  &{\rm exp}\left(-\frac{1}{2}(\Delta\hat{\vec{\iota}}_k-\Delta\vec{\iota}_k)^{T}[\Sigma_k(S^2)]^{-1}(\Delta\hat{\vec{\iota}}_k-\Delta\vec{\iota}_k)    \right)~,
\end{split}
\end{equation}
where $N_{\rm DR}$ represents the number of Gaia data releases ($N_{\rm DR}=2$ when we use both GDR2 and GDR3), $\Sigma_k$ is the covariance provided by Gaia catalog, and $S$ is the error inflation factor. $\Delta\vec{\iota}_k$ is the five-parameter astrometric vector for catalog $k$ expressed relative to GDR3, e.g.,
\begin{equation}
\begin{split}
\Delta\vec{\iota}_k\equiv
&\,((\alpha_k-\alpha_{\rm DR3})\,{\rm cos\delta_k},\,\delta_k-\delta_{\rm DR3},\,\\
&\varpi_k-\varpi_{\rm DR3},\,\mu_{\alpha k}-\mu_{\alpha \rm DR3},\,\mu_{\delta k}-\mu_{\delta \rm DR3}),
\end{split}
\end{equation}
whereas $\Delta\hat{\vec{\iota}}_k$ is the five synthetic astrometric vector obtained by fitting the standard five-parameter astrometric model to the simulated abscissae. 
We note that the GDR2 and GDR3 astrometry are not strictly independent because their underlying observations partially overlap. Since the cross-release covariance is unavailable, we treat the two catalog solutions as independent in our likelihood. We empirically tested the impact of this approximation using systems with Gaia non-single-star orbital solutions and found no significant systematic bias in the inferred orbital inclinations (Zhao et al. in Prep).
The frame rotation between GDR2 and GDR3 is calibrated using the recommended parameters derived by \citet{Feng2024ApJS}\footnote{A Python script is publicly available at \url{https://github.com/ruiyicheng/Download_HIP_Gaia_GOST}}. This correction mitigates spurious astrometric offsets between the two Gaia catalogs and prevents overestimation of the orbital astrometric signal \citep{Feng2024ApJS,Thompson2026arXiv}.
More details can be found in the Appendix of \citet{Xiao2024MNRAS} and \citet{Xiao2026AJ}.

We adopt uniform priors for most fitting parameters (Table~\ref{tab:prior} of Appendix). 
With the total likelihood, $\mathcal{L}=\mathcal{L}_{\rm RV}\cdot\mathcal{L}_{\rm hip}\cdot\mathcal{L}_{\rm gaia}$, we finally derive the orbital solution by sampling the posterior via the parallel-tempering Markov Chain Monte Carlo (MCMC) sampler \texttt{ptemcee} \citep{Vousden2016}. 
We employ 30 temperatures, 100 walkers, and 80,000 steps per chain to generate posterior distributions for all the fitting parameters, with the first 40,000 steps being discarded as burn-in. Convergence was assessed by visually inspecting the cold chain and by computing the Gelman--Rubin statistic \citep{Gelman1992}, with all fitted parameters satisfying $\hat{R}<1.01$. We adopt the posterior median as the best-fit value and quote the 16th--84th percentile credible interval as the $1\sigma$ uncertainty.

Among the 35 systems in the parent sample, 24 do not have available Hipparcos measurements. For these systems, we therefore constrain the orbital solutions using Gaia astrometry, i.e., $\mathcal{L}=\mathcal{L}_{\rm RV}\cdot\mathcal{L}_{\rm gaia}$. 
In addition, three systems host more distant stellar companions identified by Gaia astrometry. For these systems, the inclinations of the stellar companions are derived using the \texttt{lofti\_gaia} package \citep{Pearce2020ApJ}; however, they are not included in the population analysis.

\subsection{Minimum mutual inclination}
In principle, the mutual inclination between orbits of two planets is determined by their individual inclinations ($I$) and longitudes of the ascending nodes ($\Omega$). The former denotes the angle between a planet's orbital angular-momentum vector and the observer's line of sight ($-\hat{z}$), and the latter represents the angle between the line of the ascending node and the reference direction ($+\hat{y}$). The true mutual inclination between planets can be expressed as
\begin{equation}
\cos \psi=\cos I_b \cos I_c + \sin I_b \sin I_c \cos(\Omega_b-\Omega_c).
\end{equation}
However, the longitude of the ascending node of the inner transiting planet is typically unconstrained, and only the inclination can be inferred. This suggests the orbital axis of a transiting planet can point in any random direction, yet its orbital plane remains aligned with the observer. For the outer, non-transiting companion, astrometry complements RV data in constraining its orbit, but the inclination may be degenerate between prograde ($I_c<90^\circ$) and retrograde ($I_c>90^\circ$) configurations \citep{Feng2025MNRAS}, leaving the true mutual inclination undetermined. Therefore, we derive the line-of-sight (minimum) mutual inclination instead, i.e., the absolute difference between the two orbital inclinations 
\begin{equation}
\Delta I=|I_b-I_c|. 
\end{equation}

In our case, $\Delta I=|90^\circ-I_{\rm out}|$. This quantity provides a lower bound on the true mutual inclination and is adopted throughout this work as a proxy for orbital misalignment. Although a small value of $\Delta I$ does not uniquely imply coplanarity for an individual system because the relative $\Omega$ remains unconstrained, a population concentrated toward small values of $\Delta I$ is unlikely to arise if the inner and outer orbital orientations are independent and isotropic. Therefore, while $\Delta I$ should not be interpreted as the true mutual inclination on a system-by-system basis, its population distribution provides statistical evidence that the inner and outer orbits are generally correlated and that large true mutual inclinations are uncommon (see Appendix~\ref{app:appendix_mapping}).

\begin{figure*}[ht!]
    \centering
    \includegraphics[width=0.85\textwidth]{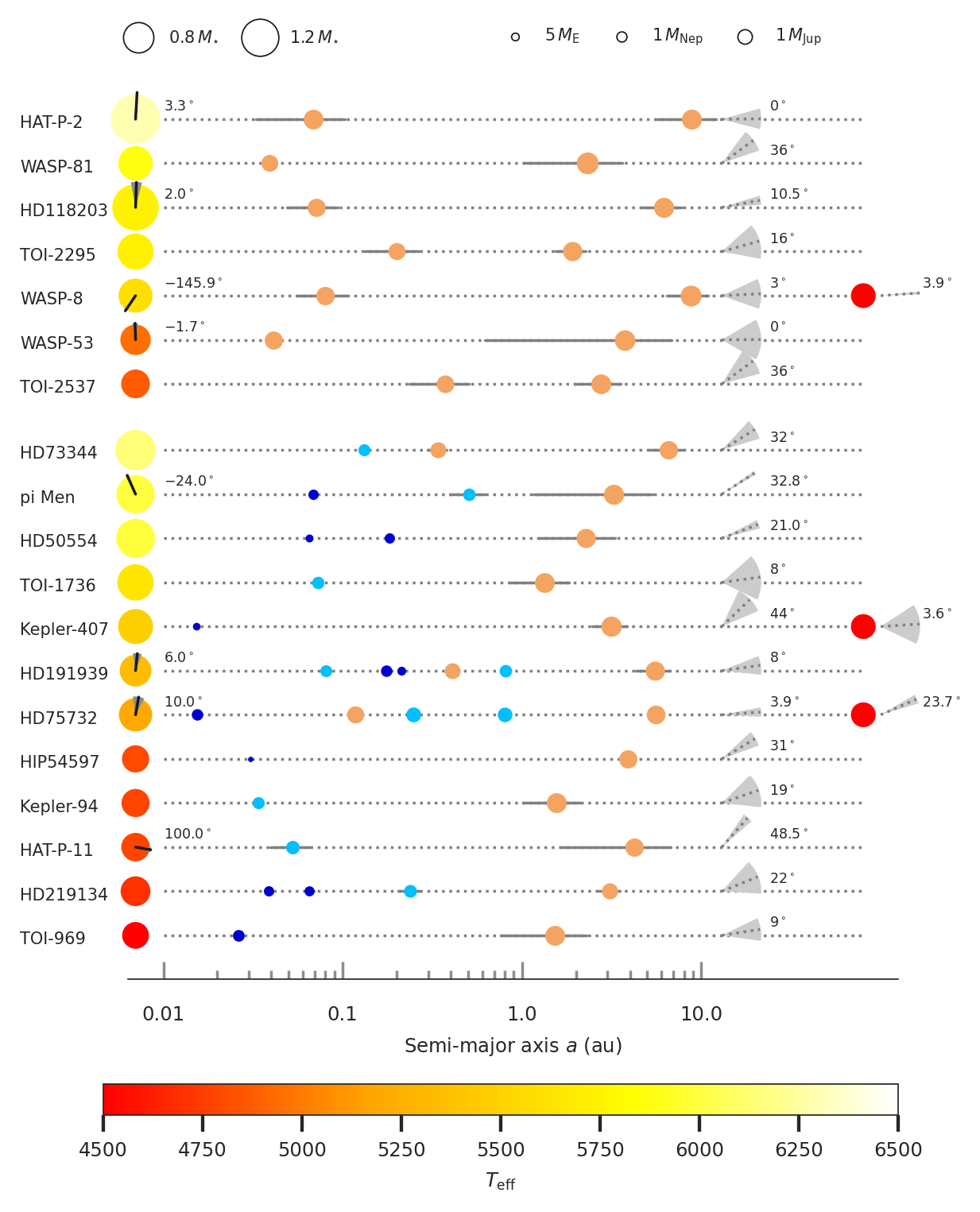}\caption{Orbital architectures of the 19 systems with well-constrained $\Delta I$ derived from the joint RV+astrometric analysis. Each row represents one planetary system. Colored circles denote planets, with the circle size proportional to the planetary (minimum) mass. Dark blue, light blue, and orange symbols correspond to super-Earths ($M_{\rm p}\lesssim10\,M_\oplus$), Neptune/Saturn-mass planets ($10\lesssim M_{\rm p}/M_\oplus\lesssim95$), and Jupiter-mass planets ($M_{\rm p}\gtrsim95\,M_\oplus$), respectively. The horizontal black bars attached to the symbols illustrate the orbital excursion between periastron and apastron. Red symbols on the right indicate outer distant stellar companions. The stellar symbols on the left show the host stars, where the symbol size scales with stellar mass and the color indicates stellar effective temperature. The black region overlaid on several host stars indicates the projected stellar obliquity angle, when available, adopted from \citet{Wang2026arXiv}. 
    The gray shaded sectors on the right illustrate the minimum mutual inclination, $\Delta I=|90^\circ-I_{\rm out}|$. The quoted values correspond to the median of the posterior distribution, while the extent of each sector indicates the $68\%$ credible interval derived from the orbital fit.
\label{fig:obirtal_architecture}}
\end{figure*}

\subsection{Orbital architecture results}
We ultimately obtain 19 systems with relatively well-constrained orbital solution ($\sigma_I<35^\circ$). 
Figure~\ref{fig:obirtal_architecture} presents the orbital architectures of these 19 systems. The systems exhibit a hierarchical architecture, with inner transiting planets located within $\sim0.01-1$ au and distant outer giant companions typically residing at several au.
In addition, the innermost transiting planets span a broad range of masses, including super-Earths, Neptune/Saturn-mass planets, and Jupiter-mass planets, whereas the outer companions reside in Jupiter and brown dwarf region.

Most systems display relatively modest value of $\Delta I$\,($\sim10^\circ-30^\circ$), consistent with only modest orbital misalignments between the inner and outer orbits. Nevertheless, several systems exhibit significantly misaligned configurations with $\Delta I\gtrsim40^\circ$, suggesting substantial dynamical excitation in a subset of systems.

HAT-P-2 might have a nearly coplanar configuration, with a $\Delta I$ consistent with zero within uncertainties. Despite hosting a massive and eccentric outer companion, the system exhibits little evidence for substantial inclination excitation, suggesting that dynamically cold architectures can survive even in the presence of distant giant perturbers (e.g., \citealt{Fabrycky2014ApJ, Winn2015ARA&A}). Systems such as pi Men and HD73344 exhibit moderate $\Delta I$ of order $\sim20^\circ$--$30^\circ$. These systems occupy the intermediate-separation regime, potentially indicating efficient secular excitation without complete disruption of the inner planetary system \citep{Naoz2016ARA&A}. Several systems exhibit strongly misaligned configurations with $\Delta I \gtrsim 40^\circ$, including Kepler-407 and HAT-P-11. Such large mutual inclinations are difficult to reconcile with quiescent disk evolution alone and instead point toward significant post-formation dynamical evolution, such as planet--planet scattering or secular perturbations induced by the outer companion (e.g., \citealt{Chatterjee2008ApJ, Naoz2016ARA&A, Lu2025ApJ}). Orbital parameters and details of each system are presented in Appendix~\ref{app:orbit_pars} and \ref{app:individual}. 

\begin{figure*}[ht!]
    \centering
    \includegraphics[width=0.65\textwidth]{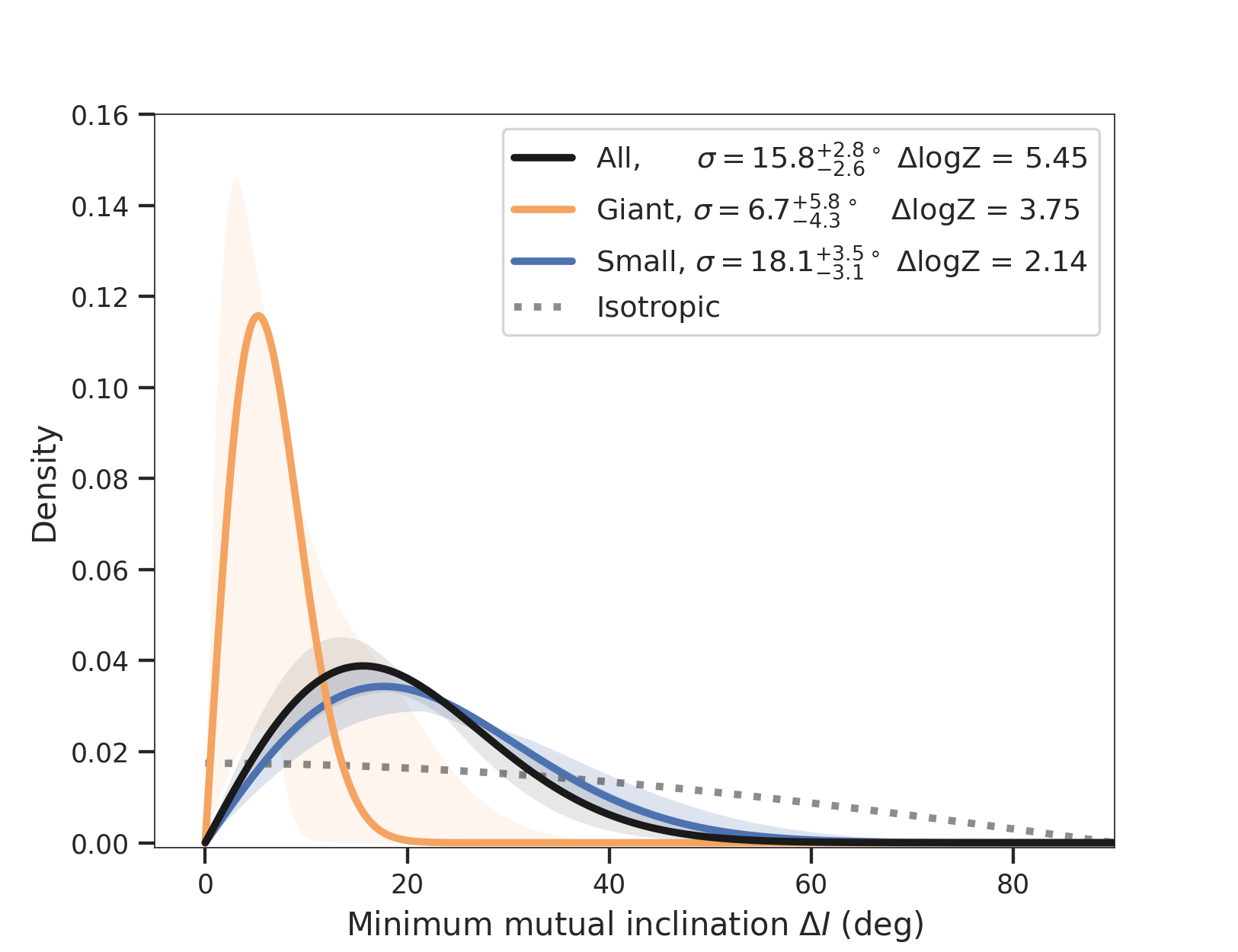}\caption{Hierarchical Bayesian inference of the $\Delta I$ distribution. The black curve shows the inferred distribution for the full sample, while the orange and blue curves correspond to systems hosting giant inner planets ($M_{\rm in}>0.3\,M_{\rm Jup}$ or $95\,M_\oplus$) and lower-mass inner planets ($M_{\rm in}<0.3\,M_{\rm Jup}$ or $95\,M_\oplus$), respectively. The gray dotted curve represents the isotropic expectation. The quoted $\Delta\log Z$ values are relative to the isotropic model. 
\label{fig:main_res}}
\end{figure*}

\section{Hierarchical Bayesian Modelling}
To determine the underlying distribution of the $\Delta I$ at the population level given the derived posteriors for individual systems, we adopt the Hierarchical Bayesian modelling (HBM). 
Given that most of our posteriors are asymmetric and non-Gaussian in shape, it is not suitable to adopt the reported values as the true values to study their distributions. 
In addition, the inclination-dependent detection efficiency of our method, which favors the recovery of relatively face-on orbits, could bias the population inference if not properly modeled.
The HBM naturally provides a flexible framework to incorporate these types of probabilistic information at multiple (individual and population) levels, and has been widely used in the exoplanet field, such as exploring the underlying eccentricity distribution.
Previous works (e.g., \citealt{Hogg2010ApJ,Bowler2020AJ}) have described an importance sampling approach to HBM with a specific application to exoplanet eccentricities. We follow their ideas, extending the HBM to minimum mutual inclinations while simultaneously incorporating the detection bias inherent in our astrometric model. 

Following previous works\citep{Fabrycky2009ApJ,Lissauer2011ApJS,Masuda2020AJ,Millholland2021AJ,Zhang2026AJ}, we assume a Rayleigh distribution for the underlying distribution of $\Delta I$. Because the range of $\Delta I$ is from 0 to $\pi/2$, we restrict and normalize the Rayleigh function to this interval.
The normalized Rayleigh distribution can be expressed as
\begin{equation}
    \mathcal{R}(x\,|\,\sigma) = \frac{\frac{x}{\sigma^2}\exp(-\frac{x^2}{2\sigma^2})}{1-\exp(-\frac{(\pi/2)^2}{2\sigma^2})}, 
\end{equation}
where $x$ denotes $\Delta I$.
The scale parameter $\sigma$, which represents the typical $\Delta I$ of our sample, determines the width of the Rayleigh distribution, i.e., a larger $\sigma$ gives a broader, flatter distribution, while a smaller $\sigma$ yields a narrower, sharper peak. The mean is $\sqrt{\pi/2}\,\sigma$ and the standard deviation is $\sqrt{2-\pi/2}\,\sigma$. 

Since the population analysis is restricted to the 19 systems satisfying our constraint criterion, we condition the likelihood on inclusion in the constrained sample. For simplicity, we denote this conditional likelihood as $\mathcal{L}(d_j\mid\sigma)$ below.
\begin{equation}
    \mathcal{L}(d_j\mid\sigma) = \frac{\int_0^{\pi/2}\mathcal{P}(d_j\mid x)\,\mathcal{R}(x\mid\sigma)\,dx}{\int_0^{\pi/2}\mathcal{P}_{{\rm det},\,j}(x)\,\mathcal{R}(x\mid\sigma)\,dx},
\end{equation}
where $j$ is the system index and $\mathcal{P}_{{\rm det},\,j}(x)$ is the individual detection rate (or the probability of entering the constrained sample) from our injection-recovery test (see Appendix~\ref{app:bias})

Applying Bayes' theorem to the individual posterior,
\begin{equation}
\mathcal{P}_{\rm post}(x\mid d_j)=\frac{\mathcal{P}(d_j\mid x)\,\pi(x)}{Z_j},
\end{equation}
where $\pi(x)$ is the interim prior and $Z_j$ is the corresponding evidence. Since $Z_j$ is independent of the population parameter $\sigma$, it can be absorbed into the proportionality constant. 
The likelihood can therefore be rewritten as
\begin{equation}
\mathcal{L}(d_j\mid\sigma)\propto\frac{\displaystyle\int_{0}^{\pi/2}\mathcal{P}_{\rm post}(x\mid d_j)\,\frac{\mathcal{R}(x\mid\sigma)}{\pi(x)}\,{\rm d}x}{
\displaystyle\int_{0}^{\pi/2}\mathcal{P}_{{\rm det},\,j}(x)\,\mathcal{R}(x\mid\sigma)\,{\rm d}x}.
\end{equation}

The numerator is evaluated by importance sampling using the $N_j$ posterior samples $x_{j,\,k}$ from the orbital fit:
\begin{equation}
\mathcal{L}(d_j\mid\sigma)\propto\frac{\displaystyle\frac{1}{N_j}\sum_{k=1}^{N_j}\frac{\mathcal{R}(x_{j,\,k}\mid\sigma)}{\pi(x_{j,\,k})}}{
\displaystyle\int_{0}^{\pi/2}\mathcal{P}_{{\rm det},\,j}(x)\,\mathcal{R}(x\mid\sigma)\,{\rm d}x}.
\end{equation}

Therefore, the total log-likelihood is
\begin{equation}
{\rm log}\,\mathcal{L}_{\rm total}=\sum\limits_{j}{\rm log}\,\mathcal{L}(d_j\mid\sigma).
\end{equation}
In our analysis, the number of constrained systems is $J=19$.

Because we adopt a prior of $\pi(I)=\sin I$ for the inclination $I$ of the outer companion (isotropic orientation), the prior for $\Delta I$ ($=|90^\circ-I_{\rm out}|$) thus can be described as 
\begin{equation}
    \pi(x) = \cos(x),\,{\rm for}\,x\in[0,\,\frac{\pi}{2}].
\end{equation}
Note that we have fixed the inclination of the inner transiting planet to $90^\circ$. Therefore, above equation is also the distribution expected under an isotropic $\Delta I$ assumption. 

We adopt a uniform prior for the scale parameters $\sigma$ ($\pi(\sigma)\sim[0,\pi/2]$) and use \texttt{dynesty} \citep{Speagle2020MNRAS} package to perform the nested sampling and calculate the logarithm of the Bayesian evidence, $\rm logZ$. 

As shown in Figure~\ref{fig:main_res}, the inferred $\Delta I$ distribution strongly favors a Rayleigh model over an isotropic configuration, with $\Delta\log Z=5.45$. The posterior gives a characteristic scale of $\sigma={15.8_{-2.6}^{+2.8}}^\circ$ ($1\sigma$), corresponding to a standard deviation of $10.4^\circ$. This result suggests that transiting systems with distant giant companions generally retain a substantial degree of coplanarity despite their large orbital separations. 
We also test a two-component Rayleigh mixture model to examine whether the sample exhibits a dichotomy similar to that reported for small planets in close binary systems \citep{Zhang2026AJ}. This more flexible model is not significantly favored over the single-Rayleigh model ($\Delta\log Z\sim0$).

We further divide the sample according to the mass of the innermost transiting planet. Systems hosting giant transiting inner planets ($M_{\rm in}>0.3\,M_{\rm Jup}$ or $95\,M_{\oplus}$) exhibit a relatively lower $\Delta I$ distribution, with $\sigma=6.7^{+5.8}_{-4.3}{}^\circ$, whereas systems hosting lower-mass transiting planets show a broader distribution with $\sigma=18.1^{+3.5}_{-3.1}{}^\circ$. The giant-planet subsample shows moderate evidence against isotropy ($\Delta \log Z=3.75$), while a weak preference over isotropy is found for the low-mass subsample ($\Delta \log Z=2.14$). 
We further compare the posterior distributions of the Rayleigh scale parameters for the two subsamples and find a $95.2\%$ probability that the giant-inner-planet systems have a smaller inclination width than the lower-mass subsample, i.e., $P(\sigma_{\rm giant}<\sigma_{\rm small})=0.952$. However, a model allowing independent Rayleigh widths, i.e., 
\begin{equation}
\log \mathcal{L}_{\rm total}=\sum\limits_{j\in \rm giant}\log \mathcal{L}_j(d_j\mid\sigma_{\rm g})+
\sum\limits_{j\in \rm small}\log \mathcal{L}_j(d_j\mid\sigma_{\rm s}),
\end{equation}
for the two subsamples is not significantly favored over a shared-width model ($\Delta\log Z=0.45$). We therefore regard the apparent difference between the two subsamples as suggestive rather than conclusive. Whether this trend reflects differences in formation history or subsequent dynamical evolution remains unclear and is discussed further in Section~\ref{sec:discussion}.

We note that the HBM explicitly accounts for the inclination-dependent astrometric recoverability within the RV parent sample, but does not attempt to reconstruct the full selection function of the underlying RV surveys. Since RV detections preferentially favor systems with larger line-of-sight velocity amplitudes, the parent sample itself is expected to be biased toward outer companions with orbital inclinations closer to edge-on configurations, corresponding to smaller $\Delta I$. Consequently, the inferred Rayleigh width should be interpreted as the effective distribution of RV-detectable transiting systems rather than the intrinsic distribution of the full underlying population.

\section{Discussion and Summary}
\label{sec:discussion}

The overall $\Delta I$ distribution inferred in this work indicates that transiting planetary systems hosting distant giant companions remain substantially more coplanar than expected for an isotropic population. The inferred Rayleigh width is $\sigma = 15.8^{+2.8}_{-2.6}$$^\circ$. This value is broadly comparable to the statistical estimate of \citet{Masuda2020AJ} ($\sigma=11.8^{+12.7}_{-5.5}$$^\circ$), who inferred the distribution of the true mutual inclination between cold Jupiters and inner super-Earths from transit occurrence statistics. Although the two quantities are not directly equivalent, both studies suggest that large mutual inclinations are uncommon.

This suggests that most outer companions are modestly tilted relative to the inner planetary system. Thus, the presence of a distant giant companion does not necessarily imply strong orbital misalignment, although the inferred $\Delta I$ width remains substantially larger than the characteristic mutual inclinations of $\sim1^\circ$--$3^\circ$ measured for compact \textit{Kepler} multi-planet systems \citep{Lissauer2011ApJS,Fang2012ApJ}. Together, these results suggest that planetary systems hosting distant giant companions retain a degree of primordial coplanarity while experiencing a greater level of dynamical excitation than typical compact multi-planet systems.

We also find tentative evidence that systems hosting giant inner planets have lower $\Delta I$ than those hosting lower-mass inner planets. The posterior probability that the giant-inner-planet subsample has a narrower $\Delta I$ distribution is $P(\sigma_{\rm giant}<\sigma_{\rm small})=0.952$. However, a model with independent $\Delta I$ widths for the two subsamples is not significantly favored over a shared-width model, indicating that the current evidence for distinct populations remains weak.

If confirmed with larger samples, such a trend could provide clues to differences in formation and dynamical evolution. For example, close-in giant planets may preferentially arise through migration channels that preserve relatively low mutual inclinations, such as disk-driven migration or coplanar high-eccentricity migration \citep{DawsonChiang2014,Petrovich2015ApJ,Dawson2018ARA&A}. The latter has been proposed as a plausible formation pathway for hot Jupiters (e.g., \citealt{Becker2017AJ, Zink2023ApJ}). Alternatively, lower-mass inner planetary systems may be more susceptible to long-term inclination excitation by distant giant companions. The present data, however, do not yet allow these possibilities to be distinguished.

The transit selection of the inner planets does not, by itself, necessarily favor systems with small inner--outer mutual inclinations. For a population with no preferred absolute orientation relative to the line of sight, inclination evolution can move systems both into and out of transit. Moreover, because our selection requires only that at least one inner planet transit, increasing the mutual inclinations within the inner system need not reduce, and can even increase, the transit probability, although dynamical loss of inner planets could have the opposite effect. A potential selection bias can nevertheless arise from the combination of RV and transit selection. RV detections preferentially favor outer companions with relatively edge-on orbits. Conditional on such an outer-orbit inclination bias, increasing the inclination difference between the inner and outer systems can reduce the probability that the inner system also satisfies the transit criterion. Thus, any bias in the sample is not a simple consequence of mutual inclination excitation, but may arise from the combined selection on the absolute orientation of the outer orbit and the transit geometry. A comprehensive forward model of the complete survey selection function is needed.

Future Gaia releases and continued long-baseline RV monitoring will enlarge the sample of systems with well-constrained orbital architectures. These measurements will provide new opportunities to investigate how distant giant companions shape planetary system evolution, including possible connections between mutual inclination, stellar obliquity, planetary multiplicity, and overall system architecture.

\begin{acknowledgments}
This work is supported by the National Key R\&D Program of China, No.2024YFA1611801 and No. 2024YFC2207700 by the National Natural Science Foundation of China (NSFC) under Grant No. 12473066 and 12573068, by the Shanghai Jiao Tong University 2030 Initiative, by SJTU-Warwick Joint Seed Fund 2024/25-Round 5. and by the China Chile Joint Research Fund (CCJRF No. 2205). CCJRF is provided by Chinese Academy of Sciences South America Center for Astronomy (CASSACA) and established by National Astronomical Observatories, Chinese Academy of Sciences (NAOC) and Chilean Astronomy Society (SOCHIAS) to support China-Chile collaborations in astronomy. The computations in this paper were run on the Siyuan-1 cluster supported by the Center for High Performance Computing at Shanghai Jiao Tong University. 
H.Y.T. appreciates the support by the EACOA/EAO Fellowship Program under the umbrella of the East Asia Core Observatories Association. 
This research has made use of the NASA Exoplanet Archive, which is operated by the California Institute of Technology, under contract with the National Aeronautics and Space Administration under the Exoplanet Exploration Program. This research has made use of data obtained from or tools provided by the portal \url{exoplanet.eu} of The Extrasolar Planets Encyclopaedia. 
This work presents results from the European Space Agency (ESA) space mission Gaia. Gaia data are being processed by the Gaia Data Processing and Analysis Consortium (DPAC). Funding for the DPAC is provided by national institutions, in particular the institutions participating in the Gaia MultiLateral Agreement (MLA). The Gaia mission website is \href{https://www.cosmos.esa.int/gaia}{https://www.cosmos.esa.int/gaia}. The Gaia archive website is \href{https://archives.esac.esa.int/gaia}{https://archives.esac.esa.int/gaia}.
\end{acknowledgments}

\begin{contribution}

All authors contributed equally to this work.


\end{contribution}

%
\facilities{Hipparcos, Gaia}

\software{astropy \citep{2013A&A...558A..33A,2018AJ....156..123A,2022ApJ...935..167A},  
          scipy \citep{2020SciPy-NMeth}, matplotlib \citep{Hunter:2007}}

\clearpage
\appendix

\section{Stellar properties}
\label{app:stellar_pars}
Table~\ref{Tab:star} summarizes the basic stellar properties and system information for the parent sample considered in this work. The columns list the system name, Hipparcos identifier, spectral type, Gaia $G$-band magnitude, stellar effective temperature $T_{\rm eff}$, surface gravity $\log g$, metallicity [Fe/H], stellar mass, distance, total number of known planets in the system, and the corresponding literature references. The stellar parameters are primarily collected from the NASA Exoplanet Archive.

\section{Summary of radial velocity measurements}
\label{app:rv_data}
Table~\ref{Tab:RV} summarizes the published radial velocity data adopted in this work. For each target and instrument, we list the spectrograph name, the total number of RV measurements $N_{\rm obs}$, the mean RV uncertainty $\langle\sigma_{\rm RV}\rangle$, the observational time baseline, and the corresponding literature references. The RV data are compiled from the discovery papers and subsequent follow-up observations, including several long-term high-precision RV surveys based on high-resolution spectroscopy.

\section{Priors for orbital parameters}
Table~\ref{tab:prior} lists the priors adopted for the orbital fit. 

\section{Mapping between the minimum and true mutual inclination distributions}
\label{app:appendix_mapping}

The HBM presented in this work directly constrains the population distribution of the minimum mutual inclination, $\Delta I$, rather than the true 3D mutual inclination, $\psi$. Since the relative longitude of ascending node is generally unconstrained, the relation between these two quantities is non-trivial. To illustrate their statistical connection at the population level, we perform a forward Monte Carlo experiment.

For a given $\psi$ distribution, we assume that it follows a Rayleigh distribution with width $\sigma_{\psi}$. To establish the geometric relation between $\psi$ and the observable $\Delta I$, we define a coordinate system in which the line of sight is along the $z$-axis and the angular-momentum vector of the inner transiting orbit lies along the $x$-axis. We describe each orbital plane by its unit angular-momentum vector, $\hat{\boldsymbol{L}}\equiv\boldsymbol{L}/|\boldsymbol{L}|$, such that only the orientation of the orbit, rather than the magnitude of its angular momentum, is relevant. For an exactly edge-on inner orbit ($I_{\rm in}=90^\circ$), we therefore have
\begin{equation}
\hat{\boldsymbol{L}}_{\rm in}=(1,0,0).
\end{equation}
The angular-momentum vector of the outer orbit is tilted from $\hat{\boldsymbol{L}}_{\rm in}$ by the true mutual inclination $\psi$. Its orientation around $\hat{\boldsymbol{L}}_{\rm in}$ is specified by an azimuthal angle $\phi$, such that
\begin{equation}
\hat{\boldsymbol{L}}_{\rm out} = (\cos\psi,\, \sin\psi\cos\phi,\, \sin\psi\sin\phi).
\end{equation}
Here, $\phi$ describes the otherwise unconstrained orientation of the mutual tilt around the inner orbital angular-momentum vector. We assume $\phi$ to be uniformly distributed over $0-2\pi$.

Since the $z$-component of $\hat{\boldsymbol{L}}_{\rm out}$ is $\cos I_{\rm out}$, the above geometry gives
\begin{equation}
\cos I_{\rm out}=\sin\psi\,\sin\phi.
\end{equation}
Using the definition $\Delta I=|90^\circ-I_{\rm out}|$, we then obtain
\begin{equation}
\Delta I = \arcsin\!\left(|\sin\psi\,\sin\phi|\right).
\end{equation}

For each adopted $\sigma_{\psi}$, we randomly draw $\psi$ from the Rayleigh distribution and $\phi$ from a uniform distribution. The resulting $\Delta I$ samples are subsequently fitted with the same Rayleigh distribution truncated to $[0,\pi/2]$ as adopted in our hierarchical Bayesian analysis, yielding the corresponding $\sigma_{\Delta I}$.

\begin{figure*}
\centering
\includegraphics[width=0.45\textwidth]{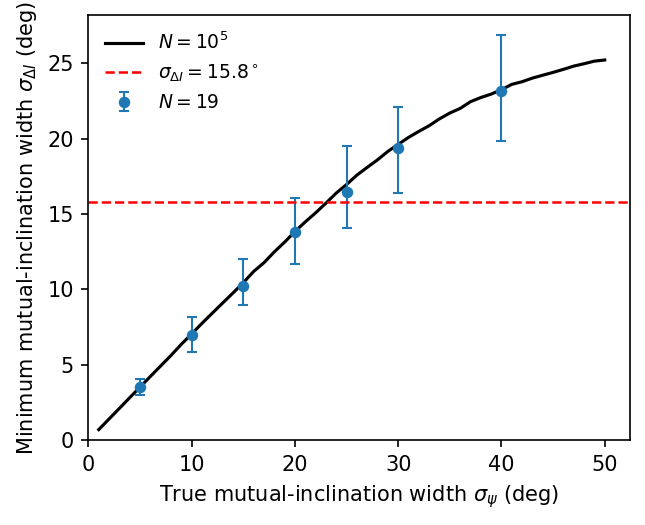}
\caption{
Forward Monte Carlo mapping between the Rayleigh width of the true mutual inclination ($\sigma_{\psi}$) and that of the minimum mutual inclination ($\sigma_{\Delta I}$). The black curve is obtained from simulations with $10^5$ systems, while the blue points show the median and $68\%$ intervals from repeated realizations with $N=19$, corresponding to the sample size of this work. The red dashed line marks the observed value, $\sigma_{\Delta I}=15.8^\circ$.
}
\label{fig:psi_deltaI_mapping}
\end{figure*}

Figure~\ref{fig:psi_deltaI_mapping} shows the recovered mapping between $\sigma_{\psi}$ and $\sigma_{\Delta I}$. The solid curve represents simulations with $10^5$ systems, while the blue points show the median and $68\%$ intervals obtained from repeated realizations with $N=19$, matching the sample size of this work. The mapping is monotonic, indicating that populations with smaller $\psi$ widths systematically produce smaller $\Delta I$. Owing to geometric projection, however, $\sigma_{\Delta I}$ is consistently smaller than $\sigma_{\psi}$, implying that $\Delta I$ provides a lower bound on the characteristic width of the true mutual-inclination distribution.

This experiment demonstrates that, although  $\Delta I$ should not be interpreted as $\psi$ for individual systems, its population distribution remains a robust statistical tracer of the underlying mutual-inclination distribution. We emphasize that this mapping is derived under the assumption of isotropically distributed azimuthal angles and neglects observational uncertainties and selection effects; therefore, it is intended solely as an illustrative geometric relation rather than a population inference.

\section{Planetary orbital parameters}
\label{app:orbit_pars}
Table~\ref{Tab:pl_orb} presents the posterior orbital parameters of the outer companions derived from the joint RV+astrometric analysis. The listed quantities include the orbital period $P$, RV semi-amplitude $K$, eccentricity $e$, argument of periastron $\omega$, longitude of ascending node $\Omega$, orbital inclination $I$, companion mass $M_{\rm p}$, semi-major axis $a$, and the inferred mutual inclination proxy $\Delta I$. The reported parameters correspond to the median posterior values, with uncertainties derived from the 16th and 84th percentiles of the posterior distributions obtained from the MCMC orbital fits. Additional figures illustrating the joint fits to the RV and astrometric data are available on \dataset[Zenodo]{https://doi.org/10.5281/zenodo.20622956}.

\section{Notes on Individual Systems}
\label{app:individual}

The individual $\Delta I$ constraints are summarized below, sorted by their inferred values. The descriptions focus on the inferred $\Delta I$, previous dynamical constraints when available, and the presence of stellar obliquity measurements or wide stellar companions.

\paragraph{HAT-P-2.}
We infer a $\Delta I$ consistent with zero, $\Delta I=0^{+15}_{-15}$$^\circ$, consistent with a nearly coplanar configuration between the inner transiting planet and the outer giant companion. The host star is also known to possess a small projected stellar obliquity, $\lambda\approx3.3$$^\circ$ \citep{Wang2026arXiv}, suggesting that the stellar spin axis is broadly aligned with the planetary orbits. Our result is therefore consistent with previous studies indicating a largely aligned orbital architecture for the system \citep{de_Beurs2023AJ_HAT-P-2}.

\paragraph{WASP-53.}
WASP-53 hosts a transiting hot Jupiter and a massive outer companion \citep{Triaud2017MNRAS_WASP-53_WASP-81_star}. We infer a minimum mutual inclination of $\Delta I=0^{+29}_{-31}$$^\circ$, consistent with a coplanar configuration within the current uncertainties. Additional astrometric constraints will be required to better determine the 3D architecture of the system.

\paragraph{WASP-8.}
WASP-8 is consistent with a nearly coplanar planet--planet configuration, with $\Delta I=3^{+23}_{-21}$$^\circ$. However, the host star exhibits a large projected stellar obliquity ($\lambda\approx-145.9$$^\circ$, \citealt{Wang2026arXiv}). The system also hosts a wide stellar companion with $\Delta I\approx3.9$$^\circ$, indicating that the planetary and stellar companion orbits may remain broadly aligned despite the large stellar spin--orbit misalignment.

\paragraph{55 Cnc (HD\,75732).}
The 55 Cnc system is consistent with a nearly coplanar architecture, with $\Delta I=3.9^{+6.9}_{-6.7}$$^\circ$. Despite hosting multiple known planets and a distant giant companion, the overall orbital configuration remains dynamically cold\citep{Nelson2014MNRAS_55Cnc}. The host star exhibits a small projected stellar obliquity ($\lambda\approx10$$^\circ$, \citealt{Wang2026arXiv}). A wide stellar companion is also present, with a moderately inclined $\Delta I$ of approximately $23.7$$^\circ$ relative to the innermost planetary plane.

\paragraph{HD\,191939.}
We measure a minimum mutual inclination of $\Delta I=8^{+12}_{-16}$$^\circ$, indicating that the outer giant companion is broadly aligned with the inner planetary system. This result is consistent with the dynamical analysis of \citet{Lubin2024AJ_HD191939}, which found that the system possesses a well-aligned and nearly coplanar architecture. The projected stellar obliquity is also small ($\lambda\approx6$$^\circ$, \citealt{Wang2026arXiv}), further supporting a largely aligned 3D configuration.

\paragraph{TOI-1736.}
TOI-1736 exhibits a small value of $\Delta I=8^{+34}_{-33}$$^\circ$. Although the inclination uncertainty remains relatively large, the posterior distribution is centered on a nearly coplanar configuration. Additional astrometric measurements will be required to better constrain the system geometry.

\paragraph{TOI-969.}
We measure a small value of $\Delta I=9^{+17}_{-16}$$^\circ$. The outer giant companion remains approximately aligned with the inner transiting planet, indicating a largely coplanar architecture.

\paragraph{HD\,118203.}
HD\,118203 exhibits a moderate minimum mutual inclination of $\Delta I=10.5^{+6.3}_{-7.4}$$^\circ$. The projected stellar obliquity is small ($\lambda\approx2$$^\circ$, \citealt{Wang2026arXiv}), indicating that the overall architecture remains broadly aligned despite modest planet--planet misalignment \citep{Zhang2024AJ_HD118203}.

\paragraph{TOI-2295.}
We measure a minimum mutual inclination of $\Delta I=16^{+21}_{-31}$$^\circ$. Additional astrometric constraints will be required to better determine its orbital architecture.

\paragraph{Kepler-94.}
Kepler-94 exhibits a value of $\Delta I=19^{+17}_{-33}$$^\circ$. The uncertainties remain substantial.

\paragraph{HD\,50554.}
HD\,50554 exhibits a minimum mutual inclination of $\Delta I=21.0^{+6.5}_{-6.5}$$^\circ$.
The inferred inclination indicates a modest departure from coplanarity between the inner transiting planet and the outer giant companion.

\paragraph{HD\,219134.}
HD\,219134 exhibits a value of $\Delta I=22^{+19}_{-33}$$^\circ$,  although the uncertainties remain relatively large. The host star is known to harbor a compact multi-planet system \citep{Motalebi2015A&A_HD219134}, and our result suggests that the outer giant companion remains broadly aligned with this architecture.

\paragraph{HIP\,54597.}
HIP\,54597 exhibits a minimum mutual inclination of $\Delta I=31^{+10}_{-13}$$^\circ$. The system therefore belongs to the relatively misaligned subset of the sample. 

\paragraph{HD\,73344.}
HD\,73344 exhibits a value of $\Delta I=32^{+13}_{-13}$$^\circ$, placing it among the strongly misaligned systems identified in this work. A recent 3D orbital analysis of the system similarly found that the outer Jupiter analog is strongly misaligned with respect to the inner planetary system and disfavors coplanar architectures \citep{Zhang2025AJ_HD73344}. Our result is therefore broadly consistent with previous evidence for a substantially tilted planetary architecture.

\paragraph{$\pi$ Men (HD\,39091).}
We infer a substantial minimum mutual inclination of $\Delta I=32.8^{+7.4}_{-6.3}$$^\circ$ for the $\pi$ Men system. Previous Hipparcos--Gaia astrometric analyses have also found that the outer giant planet is significantly inclined relative to the inner transiting super-Earth, strongly disfavoring a coplanar configuration \citep{DeRosa2020A&A_pimen,Damasso2020A&A_pimen,Xuan2020MNRAS_pimen}. The host star exhibits a projected stellar obliquity of $\lambda\approx-24$$^\circ$ \citep{Kunovac2021MNRAS_pimen}, making $\pi$ Men one of the clearest nearby examples of a 3D misaligned planetary architecture.

\paragraph{WASP-81.}
WASP-81 exhibits a value of
$\Delta I=36^{+14}_{-21}$$^\circ$, suggesting a moderate departure from coplanarity between the inner transiting planet and the outer companion.

\paragraph{TOI-2537.}
TOI-2537 exhibits a value of $\Delta I=36^{+19}_{-24}$$^\circ$, placing it among the higher-inclination systems in our sample. Although the uncertainty remains substantial, the posterior distribution favors an inclined configuration relative to a coplanar architecture.

\paragraph{Kepler-407.}
Kepler-407 exhibits a value of $\Delta I=44^{+16}_{-25}$$^\circ$, placing it among the highest-inclination systems in our sample. A wide stellar companion is present, with $\Delta I\approx3.6$$^\circ$ relative to the inner planetary plane. The contrast between the companion inclination and the inferred planet--planet mutual inclination highlights the complex 3D architecture of the system.

\paragraph{HAT-P-11.}
HAT-P-11 exhibits the largest mutual inclination in our sample, $\Delta I=48.5^{+6.0}_{-9.1}$$^\circ$, indicating a substantial misalignment between the inner transiting planet and the outer giant companion. The host star is also known to possess a large projected stellar obliquity, $\lambda\approx103$$^\circ$ \citep{Wang2026arXiv}, making HAT-P-11 one of the few systems with strong evidence for both spin--orbit and planet--planet misalignment. Our result is consistent with a highly non-coplanar 3D architecture for the system\citep{Xuan2020MNRAS_pimen,An2025AJ_HAT-P-11}.

\section{Detection Bias on minimum mutual inclination}\label{app:bias}
To evaluate the detection bias of $\Delta I$ as determined by our method, we utilize an injection-recovery technique. This approach involves synthesizing planet signals across a range of orbital elements to quantify the detection probability. We proceed under the assumption that the majority of orbital elements for both the inner and outer planets have been previously constrained through transit or radial velocity observations. Consequently, our analysis focuses exclusively on the remaining parameters: the inclination ($I$) and the longitude of the ascending node ($\Omega$) of the outer companion. We focus on the astrometric part of the model as these two parameters can be only constrained by astrometry.
For a given system, we uniformly sample $I$ in cosine space ranging from -1 to 1, and $\Omega$ ranging from 0 to $2\pi$, while fixing the other parameters (period $P$, RV amplitude $K$, eccentricity $e$, argument of periastron $\omega$, mean anomaly $M_0$ and minimum mass $M_p\sin I$) to the maximum a posteriori (MAP) values that are determined from our RV analysis.

\subsection{Injection and recovery of signals}

For 19 systems constrained with our joint RV and astrometric model, we perform injection--recovery simulations as follows.

\begin{enumerate}
\item Using the best-fit orbital solution, which is available for the 19 systems with acceptable astrometric constraints, we first compute the residuals of the one-dimensional along-scan (AL) abscissae for Hipparcos (if available), GDR2, and GDR3.

\item We then simulate synthetic AL abscissae for Hipparcos, GDR2, and GDR3 by adding a single-Keplerian astrometric signal, the linear motion of the target-system barycenter (TSB), and the parallax motion to the corresponding residuals. The combined abscissae, denoted by $w$, are treated as the mock observations.

\item We fix the astrometric parameters of the TSB,
$(\Delta\alpha_\ast^{\rm b},\Delta \delta^{\rm b},
\mu_{\alpha_\ast}^{\rm b},\mu_\delta^{\rm b},\varpi^{\rm b})$,
to the values derived from the best-fit solution. The remaining unknowns are therefore the four Thiele--Innes (TI) coefficients, $A$, $B$, $F$, and $G$. The model can be written as
\begin{equation}
\begin{split}
w=\,
&(\Delta\alpha_\ast^{\rm b}+\mu_{\alpha_\ast}^{\rm b}t)\sin\psi
+(\Delta \delta^{\rm b}+\mu_\delta^{\rm b}t)\cos\psi
+\varpi^{\rm b} f_{\varpi} \\
&+(BX+GY)\sin\psi +(AX+FY)\cos\psi ,
\end{split}
\end{equation}
where $X$ and $Y$ are the elliptical rectangular coordinates, which depend on the eccentric anomaly and eccentricity. The TI coefficients are obtained by solving this linear equation, and their uncertainties are estimated by propagating the catalogue astrometric uncertainties.

\item The TI coefficients are then converted into Campbell elements using the \texttt{nsstools} code\footnote{\url{https://gitlab.obspm.fr/gaia/nsstools}} \citep{Halbwachs2023}, from which we obtain the orbital inclination $I$ and its uncertainty. Specifically,
\begin{equation}
    I=\arccos\left(\frac{c_1}{c_2+\sqrt{c_2^2-c_1^2}}\right),
\end{equation}
where $c_1=AG-BF$ and $c_2=(A^2+B^2+F^2+G^2)/2$. Finally, we compute the minimum mutual inclination as $\Delta I=|90^\circ-I_c|$, assuming that the inner transiting planet has $I_b=90^\circ$.
\end{enumerate}


\subsection{Detection rate}

For each star, we inject 50,000 trial signals and repeat the recovery procedure described above. A signal is considered successfully recovered if it satisfies the following criteria: (1) the difference between the injected and recovered stellar reflex semi-major axes, $a_0$, is smaller than 0.3\,mas; (2) the signal-to-noise ratio of the astrometric signal satisfies $a_0/\Delta a_0>3$; and (3) the recovered inclination uncertainty is smaller than $35^\circ$. Since we adopt a $\sin I$ prior for the orbital inclination, uncertainties larger than $\sim40^\circ$ generally indicate weak or unconstrained inclination measurements.

\begin{figure*}
    \centering
    \includegraphics[width=0.45\textwidth]{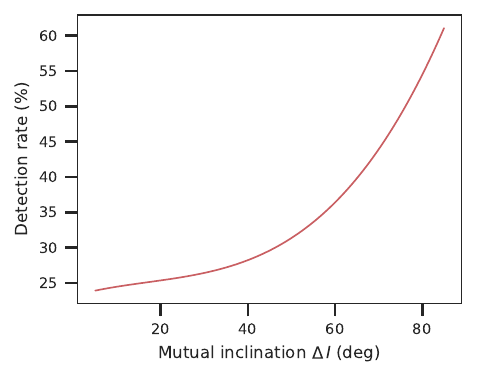}\caption{Combined detection rate as a function of $\Delta I$. 
\label{fig:detect_bias}}
\end{figure*}

We divide the minimum mutual inclination range, $\Delta I=0^\circ$--$90^\circ$, into nine uniform bins and define the detection rate in each bin as the ratio of recovered to injected signals. The detection rates for individual stars can be combined to characterize the overall sensitivity of the sample. Figure~\ref{fig:detect_bias} shows the combined detection rate as a function of $\Delta I$. Our astrometric model is preferentially sensitive to large $\Delta I$, corresponding to nearly face-on orbits ($I\sim0^\circ$ or $180^\circ$). This selection effect must therefore be incorporated into the hierarchical Bayesian inference. In practice, we use the individual detection rate of each system in the population-level analysis.

\clearpage
\begin{table}
\centering
\caption{Stellar properties for Our Sample}\label{Tab:star}
\resizebox{\textwidth}{!}{
\begin{tabular}{lcccccccccc}
\hline \hline
 Name & Hip ID  &Sp Type &$G_{\rm gaia}$ &$T_{\rm eff}$ & ${\rm log}g$ & [Fe/H] & Mass &distance &$N_{\rm planet} $ &Refs\\
 &&& (mag) &(K) & (cgs) & (dex)& ($M_\odot$)& (pc) & &\\
\hline
HAT-P-11&97657&K4V&9.15&$4778\pm113$&$4.56\pm0.09$&$0.30\pm0.05$&$0.76\pm0.04$&$37.76\pm0.03$&2&1, 2\\
HAT-P-13&---&G4&10.42&$5653\pm90$&$4.13\pm0.04$&$0.41\pm0.08$&$1.22\pm0.05$&$246.8\pm2$&2&1, 3\\
HAT-P-17&---&G0&10.27&$5246\pm80$&$4.53\pm0.02$&$0.02\pm0.09$&$0.99\pm0.15$&$92.4\pm0.5$&2&1, 4\\
HAT-P-2&80076&F8&8.60&$6290\pm60$&$4.092\pm0.074$&$0.14\pm0.08$&$1.34\pm0.04$&$127.8\pm0.4$&2&1, 5\\
HAT-P-44&---&K0V&12.97&$5295\pm100$&$4.46\pm0.06$&$0.33\pm0.1$&$0.942\pm0.041$&$347.8\pm2$&2&1, 6\\
HD118203&66192&G0V&7.89&$5742\pm93$&$3.92\pm0.02$&$0.29\pm0.07$&$1.25\pm0.04$&$92.3\pm0.2$&2&1, 7\\
HD137496&---&G5&9.77&$5799\pm61$&$4.05\pm0.1$&$-0.03\pm0.04$&$1.04\pm0.02$&$155.3\pm2$&2&1, 88\\
HD191939&99175&G9V&8.77&$5348\pm100$&$4.3\pm0.1$&$-0.15\pm0.06$&$0.85\pm0.04$&$53.61\pm0.07$&6&1, 9\\
HD219134&114622&K3V&5.24&$4699\pm16$&$4.567\pm0.018$&$0.11\pm0.04$&$0.8\pm0.09$&$6.531\pm0.004$&4&1, 10\\
HD50554&33212&F8V&6.70&$5987\pm70$&$4.38\pm0.06$&$-0.04\pm0.03$&$1.04\pm0.04$&$31.17\pm0.06$&3&1, 11\\
HD73344&42403&F6V&6.75&$6148\pm100$&$4.372\pm0.013$&$0.14\pm0.03$&$1.08\pm0.04$&$35.26\pm0.05$&3&1, 12\\
HD75732&43587&G8V&5.73&$5198\pm31$&$4.3\pm0.05$&$0.39\pm0.1$&$0.9\pm0.1$&$12.59\pm0.01$&5&1, 13\\
HD80653&---&G2&9.31&$5959\pm61$&$4.338\pm0.033$&$0.26\pm0.07$&$1.18\pm0.04$&$109.9\pm0.8$&2&1, 14\\
HD86226&48739&G1V&7.77&$5863\pm88$&$4.4\pm0.029$&$0.02\pm0.06$&$1.019\pm0.061$&$45.7\pm0.1$&2&1, 15\\
HIP54597&54597&K5V&9.48&$4799\pm90$&$4.43\pm0.18$&$-0.22\pm0.06$&$0.73\pm0.04$&$39.74\pm0.07$&2&1, 16\\
K2-73&---&G1V&11.71&$5867\pm100$&$4.39\pm0.081$&$0.04\pm0.04$&$1.02\pm0.04$&$269.2\pm3$&2&1, 17\\
KELT-6&---&F8&10.19&$6102\pm43$&$4.07\pm0.06$&$-0.27\pm0.06$&$1.42\pm0.46$&$240.7\pm2$&2&1, 4\\
Kepler-129&---&G4V&11.64&$5770\pm83$&$4.09\pm0.06$&$0.29\pm0.1$&$1.178\pm0.021$&$408.8\pm4$&3&1, 18\\
Kepler-407&---&G7&12.51&$5487\pm100$&$4.3\pm0.1$&$0.35\pm0.05$&$0.94\pm0.05$&$338.4\pm3$&2&1, 19\\
Kepler-454A&---&G4V&11.44&$5687\pm49$&$4.35\pm0.07$&$0.32\pm0.08$&$1.03\pm0.04$&$230.9\pm1$&3&1, 14\\
Kepler-88&---&G8IV&13.10&$5471\pm50$&$4.528\pm0.022$&$0.20\pm0.08$&$0.99\pm0.023$&$376.9\pm2$&3&1, 20\\
Kepler-94&---&M0V&12.86&$4781\pm98$&$4.59\pm0.04$&$0.34\pm0.07$&$0.75\pm0.09$&$191.9\pm0.8$&2&1, 21\\
TIC139270665&---&G2&10.18&$5844\pm84$&$4.439\pm0.032$&$0.16\pm0.06$&$1.035\pm0.052$&$189.9\pm2$&2&1, 22\\
TOI-1736&---&G5&8.78&$5636\pm86$&$4.142\pm0.024$&$0.15\pm0.06$&$0.98\pm0.04$&$88.9\pm0.3$&2&1, 23\\
TOI-2295&---&G3V&9.46&$5730\pm140$&$4.176\pm0.042$&$0.32\pm0.04$&$0.97\pm0.04$&$126.3\pm0.4$&2&1, 24\\
TOI-2537&---&K3V&12.73&$4870\pm150$&$4.551\pm0.049$&$0.08\pm0.08$&$0.77\pm0.04$&$185.0\pm2$&2&1, 24\\
TOI-4010&---&K3V&12.12&$4960\pm36$&$4.54\pm0.02$&$0.37\pm0.07$&$0.88\pm0.03$&$177.5\pm1$&4&1, 25\\
TOI-969&---&K5&11.27&$4435\pm80$&$4.55\pm0.06$&$0.18\pm0.02$&$0.72\pm0.04$&$77.3\pm0.3$&2&1, 26\\
WASP-4&---&G7V&12.32&$5500\pm150$&$4.45\pm0.02$&$-0.05\pm0.04$&$0.89\pm0.01$&$267.2\pm4$&2&1, 27\\
WASP-41&---&G8V&11.28&$5545\pm33$&$4.53\pm0.05$&$0.06\pm0.02$&$0.81\pm0.15$&$163.4\pm1$&2&1, 4\\
WASP-53&---&K3V&12.35&$4953\pm60$&$4.553\pm0.019$&$0.22\pm0.11$&$0.8\pm0.1$&$201.3\pm1$&2&1, 28\\
WASP-8&---&G6&9.61&$5600\pm80$&$4.5\pm0.1$&$0.17\pm0.07$&$0.92\pm0.05$&$90.0\pm0.4$&2&1, 4\\
WASP-81&---&G1V&12.33&$5870\pm120$&$4.258\pm0.022$&$-0.36\pm0.14$&$0.93\pm0.07$&$398.5\pm7$&2&1, 28\\
WASP-132&---&K4V&11.75&$4686\pm99$&$4.56\pm0.03$&$0.15\pm0.05$&$0.789\pm0.039$&$122.9\pm0.6$&3&1, 29\\
pi Men&26394&G0V&5.51&$5998\pm62$&$4.43\pm0.10$&$0.09\pm0.04$&$1.03\pm0.04$&$18.27\pm0.02$&3&1, 30\\
\hline
\end{tabular}
}
\begin{minipage}{\textwidth}
\footnotesize
\textbf{References—}
(1) \citet{GaiaCollaboration2023}; (2) \citet{Stassun2019AJ_HAT-P-11_star}; (3) \citet{Bakos2009ApJ_HAT_P-13_star}; (4) \citet{Stassun2017AJ_HAT-P-17_star}; (5) \citet{Southworth2010MNRAS_HAT-P-2_star}; (6) \citet{Hartman2014AJ_HAT-P-44_star}; (7) \citet{Zhang2024AJ_HD118203}; (8) \citet{Azevedo2022A&A_HD137496_star}; (9) \citet{Lubin2024AJ_HD191939}; (10) \citet{Gillon2017NatAs_HD219134}; (11) \citet{Liu2014RAA_HD50554}; (12) \citet{Zhang2025AJ_HD73344}; (13) \citet{Moutou2026A&A_55Cnc_star}; (14) \citet{Bonomo2023A&A_HD80653_star}; (15) \citet{Teske2020AJ_HD86226_star}; (16) \citet{Frensch2023A&A_HIP54597_star}; (17) \citet{Howard2025ApJS_K2-73_star}; (18) \citet{Zhang2021AJ_Kepler-129_star}; (19) \citet{Brinkman2025AJ_Kepler-407_star}; (20) \citet{Nesvorny2013ApJ_Kepler-88_star}; (21) \citet{Marcy2014ApJS_Kepler-94_star}; (22) \citet{Peluso2024AJ_TIC139270665_star}; (23) \citet{MacDougall2023AJ_TOI-1736_star}; (24) \citet{Heidari2025A&A_TOI-2295_star}; (25) \citet{Kunimoto2023AJ_TOI-4010_star}; (26) \citet{Lillo-Box2023A&A_TOI-969_star}; (27) \citet{Triaud2010A&A_WASP-4_star}; (28) \citet{Triaud2017MNRAS_WASP-53_WASP-81_star}; (29) \citet{Grieves2025A&A_WASP-132_star}; (30) \citet{Damasso2020A&A_pimen}. 
\end{minipage}
\end{table}

\clearpage
\begin{table*}
\centering
\caption{Published RV Data for Our Sample}\label{Tab:RV}
\resizebox{\textwidth}{!}{
\begin{tabular}{llccccllcccc}
\hline \hline
 Name & Instrument &$N_{\rm obs} $ &$\langle \sigma_{\rm RV}\rangle$ &Time span &Refs & Name & Instrument &$N_{\rm obs} $ &$\langle \sigma_{\rm RV}\rangle$ &Time span &Refs\\
 &&& (${\rm m\ s}^{-1}$) &(days) & &&&& (${\rm m\ s}^{-1}$) &(days) &\\
\hline
HAT-P-11&HIRES&261&1.3&6151&1&&HARPSpre&39&1.2&&\\
HAT-P-13&HIRES&66&1.6&2126&2&K2-73&HIRES&60&2.1&1016&26\\
HAT-P-17&HARPSN&25&2.0&2956&3, 4&KELT-6&HARPN&45&4.1&567&27\\
&KECK&42&1.8&&&&TRES&22&21.2&&\\
HAT-P-2&HIRES&72&7.3&5110&5&Kepler-129&Keck&55&2.3&3266&28\\
&post-HARPN&13&2.2&&&Kepler-407&HIRES&97&2.0&3816&29\\
&pre-HARPN&18&7.3&&&Kepler-454A&HARPSN&111&2.5&4403&30\\
HAT-P-44&FLWO&22&4.4&509&6&&HIRES&50&1.6&&\\
HD118203&ELODIE&43&14.8&7336&2,7,8,9,10&Kepler-88&HIRES&44&2.6&2438&31\\
&HET&51&8.6&&&&SOPHIE&11&11.4&&\\
&HIRES&26&1.4&&&Kepler-94&HIRES&39&2.1&4367&32\\
&KPF&76&1.3&&&TIC139270665&APF&59&8.3&457&33\\
&TNG&18&1.9&&&TOI-1736&SOPHIE&152&2.2&933&34\\
HD137496&CORALIE&30&13.2&903&11&TOI-2295&SOPHIE&44&2.9&1114&35\\
&HARPS&142&2.0&&&TOI-2537&FEROS&19&11.6&1624&35\\
HD191939&APF&254&3.5&1693&12,13&&HARPS&21&14.9&&\\
&CARMENES&138&3.0&&&&SOPHIE&46&19.6&&\\
&HARPSN&42&1.2&&&TOI-4010&HARPSN&112&5.0&544&36\\
&HIRES&111&1.2&&&TOI-969&CORALIE&21&0.0&1522&37\\
HD219134&APF&221&1.3&9189&14,15&&HARPS&67&0.0&&\\
&KECK&488&1.1&&&&PFS&10&0.0&&\\
&LICK13&26&1.9&&&WASP-4&CORALIE&45&21.2&6223&38\\
&LICK6&15&2.8&&&&HARPSpost&8&6.3&&\\
&LICK8&24&7.2&&&&HARPSpre&19&2.6&&\\
HD50554&APF&3&0.8&6884&16&&KECK&10&2.9&&\\
&KECK&43&1.7&&&WASP-41&CORALIE&100&10.0&2664&21\\
&LICK13&5&7.0&&&&HARPS&65&6.0&&\\
&LICK6&19&6.0&&&&HARPSpost&31&2.7&&\\
&LICK8&23&4.0&&&&HARPSpre&68&4.4&&\\
HD73344&APF&168&3.1&9714&17,18&WASP-53&C07&73&20.6&2178&39\\
&ELODIE&73&12.2&&&&C14&25&30.0&&\\
&HIRES&40&1.1&&&&HARPS&83&6.6&&\\
&Lick&23&6.1&&&WASP-8&HARPSpre&101&1.6&1865&40,41\\
&SOPHIE&137&1.2&&&&HIRES&9&1.7&&\\
HD75732&APF&90&1.2&10450&16,19,20&WASP-81&C07&53&25.0&1848&39\\
&ELODIE&61&7.4&&&&C14&14&41.7&&\\
&HARPN&27&0.5&&&&HARPS&32&13.6&&\\
&HARPSpre&88&0.4&&&WASP-132&COR07&22&18.6&3443&42\\
&HRS&119&5.7&&&&COR14&51&27.7&&\\
&KECK&659&1.1&&&&HARPSpost&15&2.8&&\\
&LICK13&60&2.0&&&&HARPSpost2&48&2.3&&\\
&LICK6&54&2.3&&&&HARPSpre&7&2.0&&\\
&LICK8&176&2.1&&&&PFS&46&1.9&&\\
HD80653&HARPSN&208&1.7&1068&21&pi Men&AAT&77&2.1&8831&43,44,45,46\\
HD86226&COR07&51&3.9&9548&22,23,24&&C07&11&3.6&&\\
&COR14&25&4.1&&&&C14&32&3.0&&\\
&COR98&12&4.5&&&&C98&10&4.6&&\\
&MIKE&13&3.9&&&&ESPRESSO&275&0.3&&\\
&PFS&349&1.5&&&&HARPSpost&417&0.9&&\\
HIP54597&HARPSpost&18&1.9&5157&25&&HARPSpre&128&1.1&&\\
&&&&&&&PFS&397&1.1&&\\
\hline
\end{tabular}
}
\begin{minipage}{\textwidth}
\footnotesize
\textbf{References—}
(1) \citet{Yee2024RNAAS_HAT-P-11_RV}; (2) \citet{Butler2017AJ_HIRES}; (3) \citet{Tal-Or2019MNRAS_HIRES}; (4) \citet{Bonomo2017A&A_HAT-P-17_HARPSN}; (5) \citet{de_Beurs2023AJ_HAT-P-2}; (6) \citet{Hartman2014AJ_HAT-P-44_star}; (7) \citet{da_Silva2006A&A_HD118203_ELODIE}; (8) \citet{Wittenmyer2009ApJS_HD118203_HET}; (9) \citet{Zhang2024AJ_HD118203}; (10) \citet{Maciejewski2024A&A_HD118203_TNG}; (11) \citet{Azevedo2022A&A_HD137496_star}; (12) \citet{Lubin2024AJ_HD191939}; (13) \citet{Polanski2024ApJS_HD191939_HIRES}; (14) \citet{Vogt2015ApJ_HD219134_HIRES_APF}; (15) \citet{Johnson2016ApJ_HD219134_LICK}: (16) \citet{Rosenthal2021ApJS_HD50554}; (17) \citet{Sulis2024A&A_HD73344_RV}; (18) \citet{Zhang2025AJ_HD73344}; (19) \citet{Bourrier2018A&A_55Cnc_HARPS}; (20) \citet{Naef2004A&A_55Cnc_ELODIE}; (21) \citet{Bonomo2023A&A_HD80653_star}; (22) \citet{Teske2020AJ_HD86226_star}; (23) \citet{Arriagada2010ApJ_HD86226}; (24) \citet{Marmier2013A&A_HD86226}; (25) \citet{Frensch2023A&A_HIP54597_star}; (26) \citet{Howard2025ApJS_K2-73_star}; (27) \citet{Damasso2015A&A_KELT-6}; (28) \citet{Weiss2024ApJS_Kepler-129}; (29) \citet{Brinkman2025AJ_Kepler-407_star}; (30) \citet{Gettel2016ApJ_Kepler-454A}; (31) \citet{Weiss2020AJ_Kepler-88}; (32) \citet{Marcy2014ApJS_Kepler-94_RV}; (33) \citet{Peluso2024AJ_TIC139270665_star}; (34) \citet{MacDougall2023AJ_TOI-1736_star}; (35) \citet{Heidari2025A&A_TOI-2295_star}; (36) \citet{Kunimoto2023AJ_TOI-4010_star}; (37) \citet{Lillo-Box2023A&A_TOI-969_star}; (38) \citet{Turner2022AJ_WASP-4_RV}; (39) \citet{Triaud2017MNRAS_WASP-53_WASP-81_star}; (40) \citet{Knutson2014ApJ_WASP-8_HIRES}; (41) \citet{Trifonov2020_HARPS}; (42) \citet{Grieves2025A&A_WASP-132_star}; (43) \citet{Damasso2020A&A_pimen}; (44) \citet{Gandolfi2018A&A_pimen_RV}; (45) \citet{Hatzes2022AJ_pimen}; (46) \citet{Feng2022ApJS}.
\end{minipage}
\end{table*}

\begin{table*}
     \centering
     \caption{Priors of fitting parameters}
     \begin{tabular*}{\textwidth}{llc}
         \hline\hline
Parameter & Description  & Prior \\
\hline
\textit{Orbital parameter}\\
$P$ (day)&Orbital period& Log-$\mathcal{U}(-1,16)^{[1]}$\\
$K$ (m s$^{-1}$)&RV semi-amplitude& $\mathcal{U}(10^{-6},10^{6})^{[2]}$\\
$e$&Eccentricity& $\mathcal{U}(0,1)$\\
$\omega$ (radian)&Argument of periapsis$^{[5]}$& $\mathcal{U}(0,2\pi)$\\
$M_0$ (radian) &Mean anomaly at reference epoch& $\mathcal{U}(0,2\pi)$\\
$I$ (radian)&Inclination& Cos$i$-$\mathcal{U}(-1,1)^{[3]}$\\
$\Omega$ (radian)&Longitude of ascending node& $\mathcal{U}(0,2\pi)$\\\hline
\textit{barycentre parameter}\\
$\Delta \alpha*$ (mas) &$\alpha*$ offset& $\mathcal{U}(-10^{6},10^{6})$\\
$\Delta \delta$ (mas) &$\delta$ offset& $\mathcal{U}(-10^{6},10^{6})$\\
$\Delta \mu_{\alpha*}$ (mas\,yr$^{-1}$) &$\mu_{\alpha*}$ offset & $\mathcal{U}(-10^{6},10^{6})$\\
$\Delta \mu_\delta$ (mas\,yr$^{-1}$)&$\mu_\delta$ offset& $\mathcal{U}(-10^{6},10^{6})$\\
$\Delta \varpi$ (mas) &$\varpi$ offset& $\mathcal{U}(-10^{6},10^{6})$\\\hline
\textit{Instrumental parameter}\\
$J^{\rm ins}$ (m\,s$^{-1}$)&RV jitter for ins&$\mathcal{U}(0,10^{6})$\\
$J^{\rm hip}$ (mas) &Jitter for hipparcos IAD&$\mathcal{U}(0,10^{6})$\\
$S^{\rm gaia}$ &Error inflation factor for gaia&$\mathcal{N}(1,0.1^2)^{[4]}$\\
         \hline
\end{tabular*}
\begin{minipage}{\textwidth}
\footnotesize
\textbf{Note—}
     [1] Log-$\mathcal{U}$($x_1$\ , \ $x_2$) stands for a log-uniform distribution between $x_1$ and $x_2$.
     
     [2] $\mathcal{U}$($x_1$\ , \ $x_2$) stands for a uniform distribution between $x_1$ and $x_2$.
    
     [3] Cos-$\mathcal{U}$($x_1$\ , \ $x_2$) stands for a uniform distribution between $x_1$ and $x_2$ in cosine space.

     [4] $\mathcal{N}(\mu, \sigma^2)$ is the Gaussian distribution with mean $\mu$ and standard deviation $\sigma$.

     [5] The argument of periastron of the stellar reflex motion, differing by $\pi$ with planetary orbit, i.e., $\omega_{\rm p}=\omega+\pi$.
\end{minipage}     
\label{tab:prior}
\end{table*}

\begin{table*}
\centering
\caption{Orbital parameters for the outer companion of our Sample}\label{Tab:pl_orb}
\resizebox{\textwidth}{!}{
\begin{tabular}{lccccccccc}
\hline \hline
 Name & $P$  &$K$ &$e$ &$\omega$ & $\Omega$ & $I$ & $M_p$ &$a$ &$\Delta I$\\
 &(days)&(${\rm m\ s}^{-1}$)&  &(deg) & (deg) & (deg)& ($M_{\rm Jup}$)& (au) & (deg)\\
\hline
HAT-P-11 c&${3632}_{-159}^{+20}$&${31.1}_{-1.0}^{+1.0}$&${0.608}_{-0.032}^{+0.029}$&${138.7}_{-3.1}^{+3.6}$&${228}_{-10}^{+11}$&${138.5}_{-9.1}^{+6.0}$&${2.36}_{-0.32}^{+0.33}$&${4.214}_{-0.11}^{+0.084}$&${48.5}_{-9.1}^{+6.0}$\\
HAT-P-2 c&${8225}_{-1458}^{+2441}$&${84}_{-10}^{+20}$&${0.36}_{-0.12}^{+0.13}$&${42}_{-36}^{+30}$&${160}_{-61}^{+83}$&${90}_{-15}^{+15}$&${9.9}_{-1.8}^{+3.1}$&${8.8}_{-1.1}^{+1.7}$&${0}_{-15}^{+15}$\\
HD 118203 c&${4997}_{-99}^{+92}$&${115.9}_{-5.1}^{+4.9}$&${0.261}_{-0.027}^{+0.028}$&${174.38}_{-0.32}^{+0.28}$&${199}_{-18}^{+17}$&${100.5}_{-7.4}^{+6.3}$&${11.25}_{-0.56}^{+0.57}$&${6.19}_{-0.10}^{+0.10}$&${10.5}_{-7.4}^{+6.3}$\\
HD 191939 f&${5109}_{-962}^{+764}$&${62.0}_{-7.1}^{+6.2}$&${0.213}_{-0.064}^{+0.069}$&${-22.9}_{-8.2}^{+6.2}$&${190}_{-125}^{+55}$&${98}_{-16}^{+12}$&${4.81}_{-0.77}^{+0.60}$&${5.50}_{-0.71}^{+0.56}$&${8}_{-16}^{+12}$\\
HD 219134 h&${2212}_{-13}^{+14}$&${6.68}_{-0.25}^{+0.26}$&${0.143}_{-0.031}^{+0.031}$&${-59}_{-15}^{+14}$&${221}_{-88}^{+44}$&${68}_{-19}^{+33}$&${0.413}_{-0.052}^{+0.083}$&${3.08}_{-0.13}^{+0.12}$&${22}_{-33}^{+19}$\\
HD 39091 b&${2088.79}_{-0.36}^{+0.37}$&${194.07}_{-0.20}^{+0.20}$&${0.64362}_{-0.00053}^{+0.00054}$&${-29.25}_{-0.15}^{+0.15}$&${271.5}_{-5.4}^{+5.4}$&${57.2}_{-2.8}^{+3.2}$&${11.39}_{-0.47}^{+0.50}$&${3.237}_{-0.042}^{+0.041}$&${32.8}_{-3.2}^{+2.8}$\\
HD 73344 d&${5849}_{-318}^{+268}$&${22.7}_{-3.3}^{+3.8}$&${0.22}_{-0.14}^{+0.16}$&${95}_{-258}^{+70}$&${260}_{-34}^{+35}$&${122}_{-18}^{+13}$&${2.48}_{-0.36}^{+0.41}$&${6.53}_{-0.25}^{+0.22}$&${32}_{-18}^{+13}$\\
55 Cnc d&${5084}_{-40}^{+40}$&${40.71}_{-0.55}^{+0.57}$&${0.0153}_{-0.0029}^{+0.0044}$&${-156.2}_{-3.0}^{+2.8}$&${124}_{-30}^{+28}$&${93.9}_{-6.7}^{+6.9}$&${3.23}_{-0.30}^{+0.29}$&${5.58}_{-0.26}^{+0.24}$&${3.9}_{-6.7}^{+6.9}$\\
55 Cnc B&---&---&${0.55}_{-0.37}^{+0.32}$&${183}_{-134}^{+138}$&${177}_{-42}^{+153}$&${67.7}_{-24}^{+7.0}$&---&${1259}_{-431}^{+826}$&${22.3}_{-7.0}^{+25}$\\
Kepler-407 c&${2096.5}_{-6.3}^{+6.5}$&${173.73}_{-0.78}^{+0.79}$&${0.2178}_{-0.0085}^{+0.0086}$&${-12.1}_{-2.5}^{+2.6}$&${106}_{-56}^{+148}$&${46}_{-16}^{+25}$&${14.4}_{-3.4}^{+6.4}$&${3.157}_{-0.053}^{+0.051}$&${44}_{-25}^{+16}$\\
Kepler-407 B&---&---&${0.58}_{-0.36}^{+0.32}$&${183}_{-129}^{+128}$&${191}_{-124}^{+71}$&${94}_{-31}^{+34}$&---&${1027}_{-366}^{+956}$&${4}_{-34}^{+31}$\\
Kepler-94 c&${815.76}_{-0.86}^{+0.82}$&${238.8}_{-3.6}^{+3.6}$&${0.350}_{-0.010}^{+0.010}$&${163.1}_{-1.5}^{+1.5}$&${63}_{-31}^{+245}$&${109}_{-33}^{+17}$&${9.5}_{-1.1}^{+1.6}$&${1.561}_{-0.065}^{+0.060}$&${19}_{-33}^{+17}$\\
TOI-1736 c&${569.98}_{-0.77}^{+0.77}$&${200.7}_{-1.2}^{+1.2}$&${0.3626}_{-0.0035}^{+0.0036}$&${164.53}_{-0.70}^{+0.72}$&${261}_{-59}^{+65}$&${82}_{-34}^{+33}$&${8.34}_{-0.70}^{+2.4}$&${1.339}_{-0.018}^{+0.018}$&${8}_{-33}^{+34}$\\
TOI-2295 c&${965.9}_{-4.2}^{+4.2}$&${105.4}_{-1.1}^{+1.1}$&${0.192}_{-0.012}^{+0.012}$&${39.4}_{-2.9}^{+2.9}$&${127}_{-49}^{+105}$&${74}_{-21}^{+31}$&${5.39}_{-0.40}^{+1.0}$&${1.898}_{-0.027}^{+0.026}$&${16}_{-31}^{+21}$\\
WASP-53 c&${2862}_{-130}^{+160}$&${475.7}_{-8.2}^{+8.8}$&${0.8333}_{-0.0071}^{+0.0073}$&${-160.1}_{-1.5}^{+1.7}$&${297}_{-51}^{+43}$&${90}_{-29}^{+31}$&${17.9}_{-2.0}^{+3.4}$&${3.73}_{-0.20}^{+0.19}$&${0}_{-31}^{+29}$\\
WASP-81 c&${1296.7}_{-8.2}^{+8.4}$&${1170.7}_{-7.4}^{+7.5}$&${0.5572}_{-0.0047}^{+0.0046}$&${-38.27}_{-0.60}^{+0.60}$&${230}_{-26}^{+27}$&${54}_{-14}^{+21}$&${64}_{-11}^{+17}$&${2.320}_{-0.065}^{+0.061}$&${36}_{-21}^{+14}$\\
TOI-2537 c&${1886}_{-130}^{+236}$&${144.7}_{-5.0}^{+5.4}$&${0.284}_{-0.051}^{+0.059}$&${-36.8}_{-10}^{+7.7}$&${149}_{-88}^{+123}$&${54}_{-19}^{+24}$&${8.9}_{-1.5}^{+3.7}$&${2.76}_{-0.14}^{+0.23}$&${36}_{-24}^{+19}$\\
TOI-969 c&${803.5}_{-4.2}^{+4.5}$&${296}_{-10}^{+10}$&${0.502}_{-0.012}^{+0.011}$&${-175.952}_{-0.045}^{+0.047}$&${111}_{-31}^{+64}$&${99}_{-16}^{+17}$&${9.86}_{-0.68}^{+1.1}$&${1.522}_{-0.030}^{+0.030}$&${9}_{-16}^{+17}$\\
WASP-8 c&${9608}_{-2173}^{+2488}$&${251}_{-65}^{+114}$&${0.26}_{-0.11}^{+0.13}$&${-5.6}_{-1.8}^{+1.0}$&${330}_{-28}^{+22}$&${93}_{-21}^{+23}$&${26.1}_{-7.9}^{+13}$&${8.7}_{-1.4}^{+1.4}$&${3}_{-21}^{+23}$\\
WASP-8 B&---&---&${0.52}_{-0.36}^{+0.41}$&${172}_{-109}^{+150}$&${173}_{-9.1}^{+175}$&${94}_{-1.2}^{+6.5}$&---&${449}_{-162}^{+396}$&${4.0}_{-6.5}^{+1.2}$\\
HD 50554 b&${1223.1}_{-1.9}^{+2.0}$&${93.7}_{-3.3}^{+3.7}$&${0.456}_{-0.018}^{+0.019}$&${5.3}_{-2.3}^{+2.2}$&${260.5}_{-9.0}^{+10}$&${111.0}_{-6.5}^{+6.5}$&${4.88}_{-0.26}^{+0.31}$&${2.274}_{-0.029}^{+0.029}$&${21.0}_{-6.5}^{+6.5}$\\
HIP 54597 b&${3283}_{-46}^{+46}$&${33.1}_{-2.0}^{+2.0}$&${0.035}_{-0.020}^{+0.036}$&${-164.4}_{-4.8}^{+8.7}$&${17}_{-12}^{+16}$&${121}_{-13}^{+10}$&${2.32}_{-0.27}^{+0.34}$&${3.901}_{-0.083}^{+0.080}$&${31}_{-13}^{+10}$\\
\hline
\end{tabular}
}
\begin{minipage}{\textwidth}
\footnotesize
\textbf{Note—}
The orbital parameters of three wide stellar companions, 55\,Cnc\,B, Kepler-407\,B and WASP-8\,B are derived using \texttt{lofti\_gaia} package \citep{Pearce2020ApJ}.
\end{minipage}
\end{table*}

\bibliography{sample701}{}
\bibliographystyle{aasjournalv7}



\end{document}